\documentclass[cits]{PoS}

\title{Sextet QCD: slow running and the mass anomalous dimension}

\ShortTitle{Sextet QCD}

\author{\speaker{Benjamin Svetitsky}, Yigal Shamir
\\
        Raymond and Beverly Sackler School of Physics and Astronomy, Tel~Aviv University, 69978 Tel~Aviv, Israel\\
        E-mail: \email{bqs@julian.tau.ac.il, shamir@post.tau.ac.il}}

\author{Thomas DeGrand\\
        Department of Physics,
University of Colorado, Boulder, CO 80309, USA\\
        E-mail: \email{degrand@pizero.colorado.edu}}

\abstract{I report the results of Schr\"odinger functional calculations in the SU(3) gauge theory with two flavors of color sextet fermions, defined with the Wilson--clover action using nHYP fat links. While we cannot confirm the infrared fixed point seen with thin links, we find very slow evolution of the coupling constant, so slow that extraction of the mass anomalous dimension is straightforward.}

\FullConference{The XXVIII International Symposium on Lattice Field Theory, Lattice2010\\
		June 14--19, 2010\\
		Villasimius, Italy}

\def\eval#1{\left\langle #1 \right\rangle}
\def\co{{\cal O}}

\begin{document}

\section{Introduction}
We have been working on the SU(3) gauge theory with $N_f=2$ sextet fermions for three years now~\cite{Shamir:2008pb, DeGrand:2008kx, DeGrand:2009hu,DeGrand:2010na}.
The theory's two-loop beta function possesses a Banks--Zaks zero, but at a fairly strong coupling $g^2\simeq10$.
Thus it might be in the conformal window, or it might be just below and hence a candidate for walking technicolor~\cite{Sannino:2004qp,Dietrich:2006cm}.

As should be done in any systematic lattice study, I will begin by showing what we know about the phase diagram of the lattice theory~\cite{DeGrand:2008kx,DeGrand:2010na}, which is interestingly different from that of QCD.
Then I will show our results for the running coupling in the massless theory, calculated with the Schr\"odinger functional method~\cite{Shamir:2008pb, DeGrand:2010na}.
What we learn from the latter is that the coupling runs very slowly compared to the two-loop perturbative running, which in turn is much slower than what we are used to in QCD.
Unfortunately, our extensive calculations are inconclusive when it comes to the existence of a non-perturbative fixed point.
The slow running, however, makes it easy to extract the mass anomalous dimension $\gamma(g^2)$, and we can state the fairly solid conclusion that $\gamma$ does not exceed 0.6.
This is too small for a phenomenologically viable theory of walking technicolor~\cite{Chivukula:2010tn}.

\section{The phase diagram}

We define the lattice theory with the plaquette gauge action and a Wilson--clover fermion action.
In our recent work we incorporated hypercubic smearing~\cite{Hasenfratz:2007rf} into the fermion action, which did not change the qualitative features of the phase diagram (but did move these features around).
I show the phase diagram in Fig.~\ref{fig:phase}.
The finite size $L$ of our $L^4$ lattices, besides rounding out any phase transitions, introduces a scale that can be interpreted as a nonzero physical temperature.

We note the following features of the phase diagram:
\begin{figure}[b]
\begin{center}
\includegraphics*[width=.48\columnwidth]{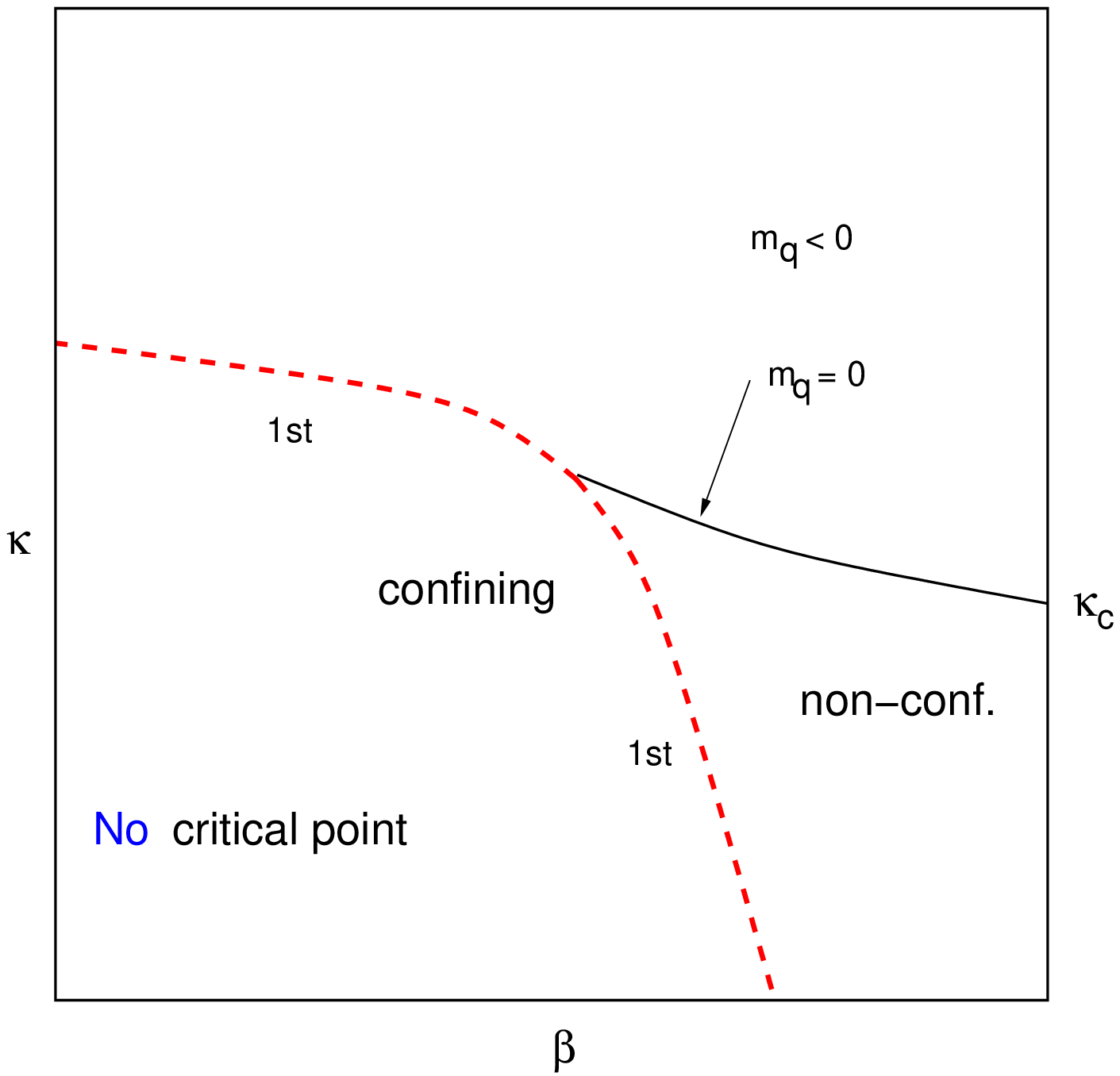}\hskip 10pt 
\includegraphics*[width=.48\columnwidth]{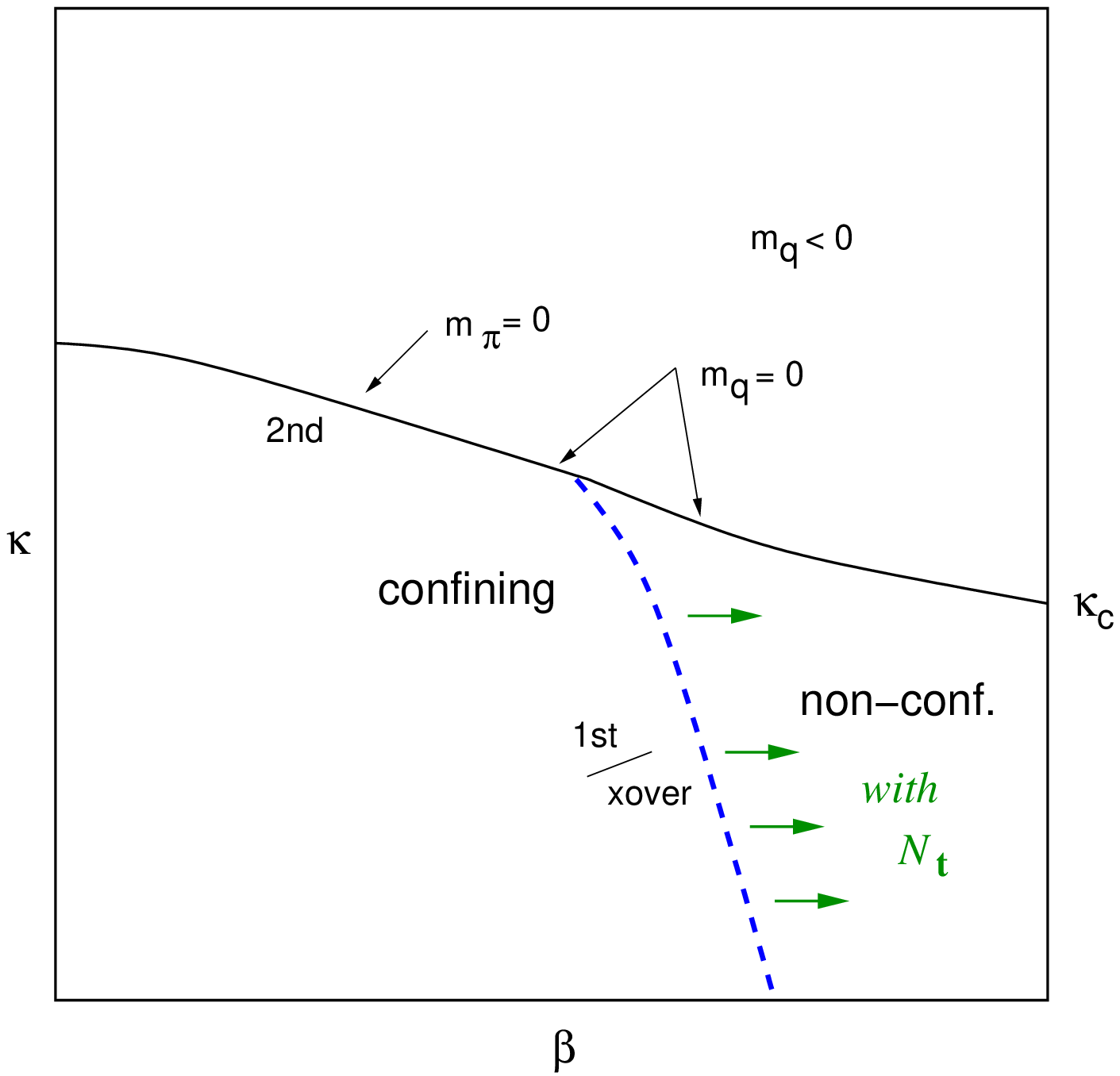}
\caption{Sketches of the phase diagram of the present theory (left) and of QCD with Wilson fermions (right) on a finite lattice.
For quantitative information see~\cite{DeGrand:2010na}.
\label{fig:phase}}
\end{center}
\end{figure}

\begin{enumerate}
\item
There is a first-order phase boundary between a strong-coupling confining phase and a weak-coupling non-confining phase.
These phases can be qualitatively distinguished by the plaquette average, by the linear piece in the potential (vs.~the absence thereof), and by the Polyakov loop.
All of these are discontinuous at the phase boundary, as is the measured AWI quark mass $m_q$.
The horizontal piece of the boundary moves little as the volume is changed, while the vertical piece behaves like a finite-temperature transition, which shifts to the right with increasing $L$.
\item
The $\kappa_c(\beta)$ curve, where $m_q=0$, exists only in the non-confining phase where it is {\em not\/} a phase transition, since there is no pion to become massless.
It intersects the phase boundary and ends there: To the left of that intersection, $m_q$ jumps at the boundary from positive to negative values without crossing zero.
This means that there is no zero-mass pion in the confining phase.
\item
There is no critical point on the phase boundary.
This situation is different from that of the SU(2) gauge theory with adjoint fermions, where the intersection point {\em is\/} a critical point~\cite{Catterall:2008qk}.
It is similar to that observed in the SU(3) theory with sufficiently many flavors of fundamental quarks~\cite{Iwasaki:2003de}.
\end{enumerate}

The phase diagram is quite different from the well-known phase diagram of QCD with Wilson fermions.
In that theory the confining phase is bounded by a second-order transition (associated with the Aoki phase) where the pion is massless.
The finite-temperature crossover moves to the right as $L$ is increased, allowing a continuum limit where confinement breaks chiral symmetry spontaneously.
It is hard to see how that might happen in the present theory.

\section{The discrete beta function}

We have previously described \cite{Shamir:2008pb,Svetitsky:2008bw} the use of the Schr\"odinger functional to calculate the running coupling on the $\kappa_c$ curve.
\begin{figure}
\begin{center}
\includegraphics*[width=.8\columnwidth]{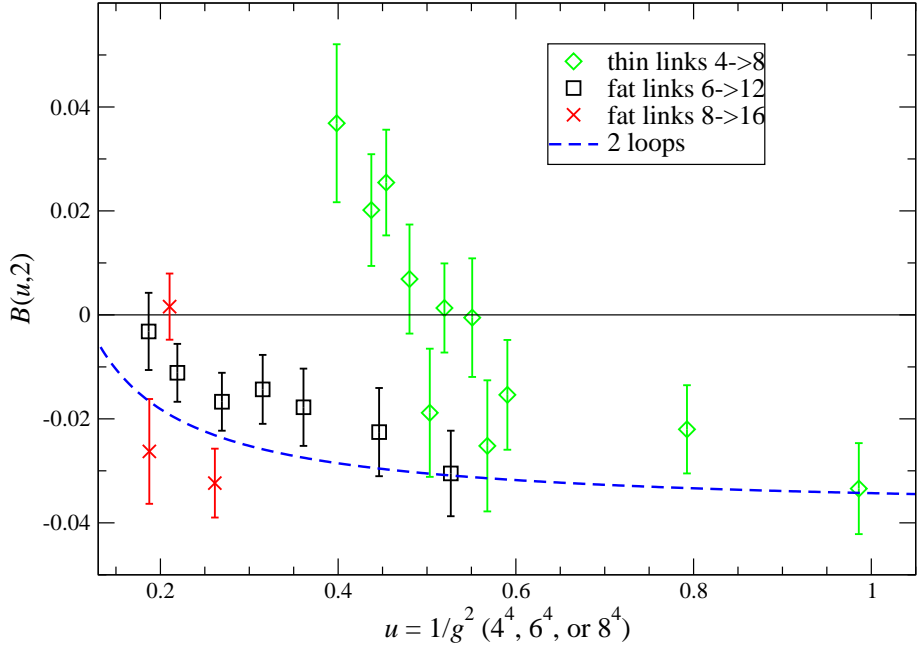}
\end{center}
\caption{Discrete beta function for scale ratio $s=2$.
\label{fig:DBF}}
\end{figure}
The discrete beta function (DBF) is defined as the shift in the inverse coupling as the scale (i.e., the lattice size) is changed from $L$ to $sL$,
\begin{equation}
B(u,s)=g^{-2}(sL)-u,\qquad{\rm where}\ u=g^{-2}(L).
\end{equation}
While our earlier data were calculated with otherwise unimproved Wilson--clover fermions (``thin links'') on small lattices, the new results shown in Fig.~\ref{fig:DBF} were obtained with nHYP-smeared fermions (``fat links'') on larger lattices.
The fat-link results {\em rule out\/} the fixed point seen with thin links, so that we may dismiss it as a small-lattice (and crude-action) artifact.
In attempting to extend the fat-link runs to stronger couplings, we run into the first-order phase transition shown above, beyond which there is no massless theory.
Thus we are left with the tantalizing Fig.~\ref{fig:DBF}, where there is no clear fixed point.
We can say unambiguously only that the coupling runs more slowly than indicated by the two-loop DBF\@.
This will be useful for discussing the anomalous dimension in the next section.

\section{Anomalous dimension}

An extended technicolor theory faces the challenge of suppressing flavor-changing neutral currents without suppressing quark masses as well.
In walking technicolor, one envisages enhancing the techniquark condensate
from the $\Lambda_{TC}$ scale to up to $\Lambda_{ETC}$ according to
\begin{equation}
\eval{\bar\Psi\Psi}_{ETC}=
\eval{\bar\Psi\Psi}_{TC}\times
\exp\left[\int_{\Lambda_{TC}}^{\Lambda_{ETC}}\frac{d\mu}{\mu}
\gamma(g^2(\mu))\right]
\end{equation}
The main feature of walking is that the coupling $g^2$ is nearly constant at an almost-fixed-point value $g_*^2$
between these scales, so this formula simplifies to
\begin{equation}
\eval{\bar\Psi\Psi}_{ETC}\simeq
\eval{\bar\Psi\Psi}_{TC}\times
\left(\frac{\Lambda_{ETC}}{\Lambda_{TC}}\right)^{\displaystyle\gamma(g_*^2)}
\end{equation}
Phenomenology~\cite{Chivukula:2010tn} requires $\gamma(g_*^2)=1$ (or {\em very} close to 1).

\begin{figure}
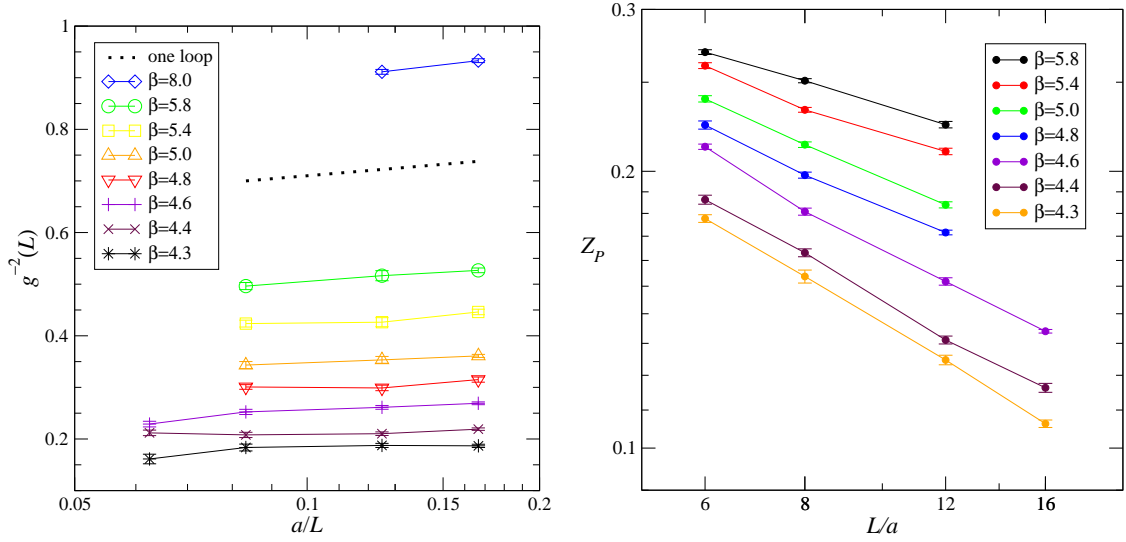

\begin{center}
\includegraphics*[width=.48\columnwidth]{1g2_vs_l.eps}\hskip 10pt
\includegraphics*[width=.48\columnwidth]{Z3.eps}
\end{center}
\caption{Running coupling (left) and pseudoscalar renormalization constant (right) as functions of lattice size $L$, for given bare coupling $\beta$. \label{fig:running}}
\end{figure}

It is straightforward to calculate $\gamma$, the anomalous dimension of $\bar\Psi\Psi$, in the framework of the Schr\"odinger functional \cite{Bursa:2009we}.
First, because the scalar operator is hard to calculate on the lattice, we note that it is related by a chiral transformation to the isovector--pseudoscalar operator $P^a=\bar\Psi(\tau^a/2)\gamma_5\Psi$, which is the pion field.
To calculate the anomalous dimension of the latter, we calculate its correlation function with a Schr\"odinger-functional wall source $\co$, propagated with zero momentum to the temporal mid-plane of the lattice according to
\begin{equation}
\left.\eval{P^b(t)\; \co^b(t'=0)}\right|_{t=L/2}=Z_P\,Z_\co\, e^{-m_\pi L/2}.
\label{eq:Z1}
\end{equation}
$Z_P$ is what we want.
To eliminate the normalization $Z_\co$ of the wall source,
we calculate the wall-to-wall propagator across the lattice,
\begin{equation}
\eval{\co^b(t=L)\; \co^b(t'=0)}=Z_\co^2\, e^{-m_\pi L}
\label{eq:Z2}
\end{equation}
Dividing (\ref{eq:Z1}) by (\ref{eq:Z2}) gives $Z_P(L)$; comparing two different lattice sizes gives the scaling relation
\begin{equation}
\frac{Z_P(L)}{Z_P(L_0)}=\left(\frac{L}{L_0}\right)^{-\gamma},
\label{eq:scaling}
\end{equation}
whence we extract $\gamma$.
In writing Eq.~(\ref{eq:scaling}) we have assumed that $\gamma$ is constant between the scales $L_0$ and $L$, which is true if the coupling doesn't run.
We are supported in this assumption by the DBF shown in Fig.~\ref{fig:DBF}.
For any fixed bare coupling $\beta$, we find that the running coupling $g^2(L)$ changes by no more than 15\% when varying $L=6\to8\to12\to16$;
see Fig.~\ref{fig:running}.

\begin{figure}
\begin{center}
\includegraphics*[width=.8\columnwidth]{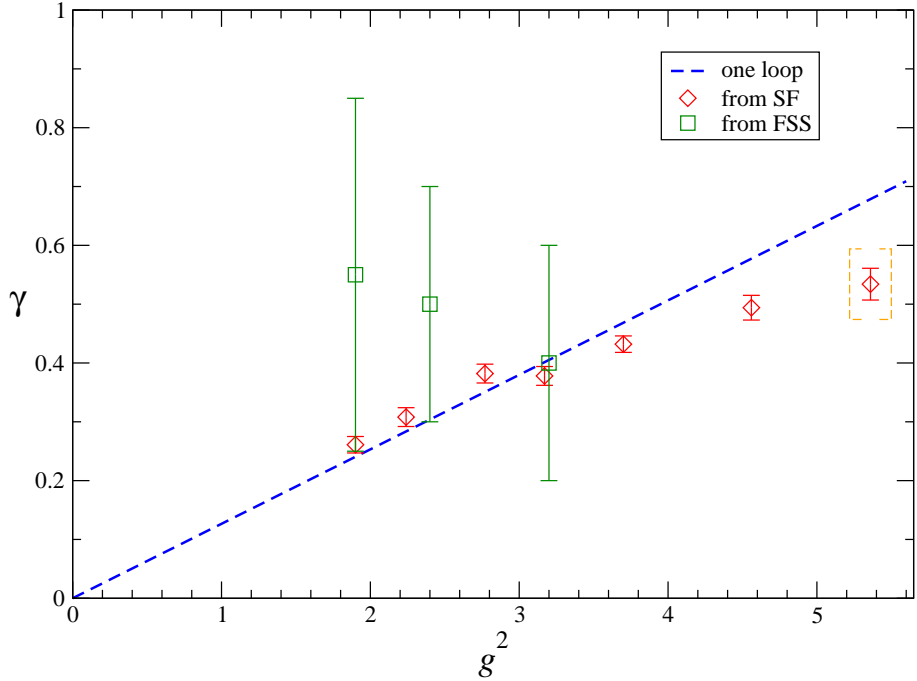}\\
\end{center}
\caption{Anomalous dimension of $\bar\Psi\Psi$, derived from $Z_P$ (red diamonds); the green squares are the results of finite-size scaling presented in Ref.~\cite{DeGrand:2009hu}.
\label{fig:gamma}}
\end{figure}

The corresponding plot of $Z_P$ shows beautiful power scaling as in Eq.~(\ref{eq:scaling}).
Linear fits to the log--log plot give, for each $\beta$, the slope $\gamma$; translating from $\beta$ to the running coupling $g^2$ gives the results plotted in Fig.~\ref{fig:gamma}.
We see that the anomalous dimension follows closely the one-loop formula
\begin{equation}
\gamma=\frac{6C_2(R)}{16\pi^2}\,g^2
\end{equation}
out to $g^2\simeq4$;
beyond that point it falls off the line and saturates below $\gamma=0.6$.

\section{Conclusions}

Since we can't tell yet whether there is an infrared fixed point in this theory, let's consider both possibilities.
Both are interesting, and each is bad news for technicolor.
\begin{itemize}
\item
{\em There is NO IR fixed point}\/:
Then the phase diagram should somehow turn into QCD when the lattice is sufficiently large.
The first-order transition, where it meets the $\kappa_c$ curve, should slide towards $\beta=\infty$ as $L\to\infty$; the discontinuity in $m_q$ (and $m_\pi$) at the intersection should tend towards zero at the same time so that the continuum limit possesses a massless (or at least a finite-mass!) pion.
The theory may confine, but its coupling still runs very slowly---maybe it walks, maybe not.
The calculation of $\gamma$ is justified by this alone, and the result $\gamma<0.6$ is bad news for technicolor.
\item
{\em There is an IR fixed point just out of reach of our calculation}\/:
Then most or all of the $\kappa_c$ line is in its catchment basin, and thus represents a conformal theory---unparticles, not technicolor.
The first-order transition will be stuck near where we found it, since it can't penetrate into a conformal phase.
Maybe a slightly different lattice action will make the fixed point accessible; maybe it's just a question of statistics.
Finally, presumably $\gamma\simeq0.6$ at the fixed point.
This may mean that the theory is deep inside the conformal window, since, according to a number of models \cite{Cohen:1988sq,Kaplan:2009kr}, $\gamma=1$ at the bottom of the conformal window.
It would be very interesting then to study the theory on the strong-coupling side of the fixed point~\cite{Pallante}, where there may or may not be found a UV-attractive fixed point, which is a critical point in the usual usage.
\end{itemize}

For studies of this theory formulated with staggered fermions, see \cite{Kogut:2010cz, Sinclair:2010be}.

This work was supported by the Israel Science Foundation
under grants no.~173/05 and no.~423/09, by the US Department of Energy, and by  
the National Science Foundation through TeraGrid resources provided by the University of Texas under grants no.~TG-PHY080042 and no.~TG-PHY090023.
Further computations were done on clusters at the University of Colorado and at Tel Aviv University.

\end{document}